\documentclass[aps,prb,showpacs,superscriptaddress,amsmath,twocolumn]{revtex4-1}

\usepackage{amsmath}
\usepackage{amssymb}
\usepackage{graphicx}

\begin{document}


\title{Transfer matrix approach for the Kerr and Faraday rotation in layered nanostructures}

\author{G\'abor Sz\'echenyi}
\affiliation{Department of Physics of Complex Systems, E\"otv\"os University, H-1117 Budapest, P\'azm\'any P{\'e}ter s{\'e}t\'any 1/A, Hungary}
\author{M\'at\'e Vigh}
\affiliation{Department of Physics of Complex Systems, E\"otv\"os University, H-1117 Budapest, P\'azm\'any P{\'e}ter s{\'e}t\'any 1/A, Hungary}
\author{Andor Korm\'anyos}
\affiliation{Department of Physics, University of Konstanz, D-78464 Konstanz, Germany}
\author{J{\'o}zsef Cserti} 
\affiliation{Department of Physics of Complex Systems, E\"otv\"os University, H-1117 Budapest, P\'azm\'any P{\'e}ter s{\'e}t\'any 1/A, Hungary}



\begin{abstract}

To study the optical rotation of the polarization of light incident on 
multilayer systems consisting of atomically thin conductors  and  dielectric multilayers we present a general method based on transfer matrices.
The transfer matrix of the atomically thin conducting layer is obtained using the Maxwell equations.
We derive expressions for the Kerr (Faraday) rotation angle and for the ellipticity 
of the reflected (transmitted) light as a function of the incident angle and polarization of the light.
The method is demonstrated by  calculating the Kerr (Faraday) angle for bilayer graphene in the quantum anomalous Hall state 
placed on the top of dielectric multilayers. The optical conductivity of the bilayer graphene 
is calculated  in the framework of a four-band model.  
 

\end{abstract}


\maketitle


\section{Introduction}
\label{intro:sec}

\begin{figure} 
\includegraphics[scale=0.35]{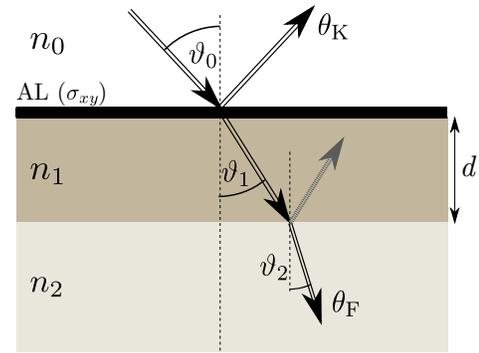}
\caption{\label{geo_layers:fig} 
Geometrical configuration of the measurement of the Kerr (Faraday) angle $\theta_\mathrm{K}$ ($\theta_\mathrm{F}$). 
An incident light with angle $\vartheta_0$ propagating in vacuum with refractive index $n_0$ 
reflected (transmitted) on an atomic layer (AL) of material 
(e.g., graphene) separated by a dielectric layer 
(e.g., $\rm{SiO}_2$) of thickness $d$ with refractive index $n_1$ from a thick substrate (e.g., $\rm{Si}$) with refractive index $n_2$.
} 
\end{figure}

Owing to the potential applications and interesting electronic properties,   
atomically thin materials have 
attracted a strong interest in recent years. A variety of two dimensional (2D) crystals, including 
graphene, boron nitride, phosphorene, several transition metal dichalcogenides 
and complex oxides, has been prepared and  studied experimentally~\cite{Novoselov_atomic_2D:cikk,Xu_Nat_Phys_2014_343,RadisavljevicB.2011,2D_Materials_2053-1583-1-2-025001,SMLL:SMLL201001628}.
Atomically thin materials are usually  fabricated and studied in multi-layer structures. 
For example, monolayer graphene placed on a substrate can hardly be observed with optical microscopy since the intensity of the reflected light is small resulting in low contrast.
However, as it was demonstrated  in Refs.~\onlinecite{Novoselov_graphene-1,Novoselov_fine-constant:ref}, 
the multilayer structure shown in Fig.~\ref{geo_layers:fig}, when a dielectric spacer of width $d$ and 
refractive index $n_1$ is placed between the substrate (with refractive index $n_2$)
and the graphene layer, can have important advantages.
Namely, by tuning the width $d$ of the $\mathrm{Si O}_\mathrm{2}$ used as spacer material, 
the intensity of the reflected light changes drastically
and consequently the visibility of the graphene flake~\cite{Blake_visibile_GR:cikk} is improved. 
Theoretically, the optical visibility of monolayer and bilayer graphene 
deposited on a Si/SiO$_2$ layer substrate was also studied in 
Ref.~\onlinecite{ISI:000248661400131} where it was shown that 
the visibility is enhanced  through a resonant transmission of light due to the spacer. 

Optical spectroscopies are powerful contact-free methods to study material properties. 
In the context of 2D materials, e.g., Zhang \textit{et al.}~\cite{PhysRevB.78.235408} and Kuzmenko \textit{et al.}~\cite{PhysRevB.80.165406} used infrared spectroscopy
to extract the tight-binding parameters in bilayer graphene by fitting the experimental reflectivity 
spectra with the optical conductivity calculated from the Kubo formula. 
If time reversal symmetry is broken, then the rotation of polarization of the transmitted (reflected) light, 
i.e., the Faraday (Kerr) effect  can be  used to deduct the off-diagonal element of the optical conductivity 
$\sigma_{xy}(\omega)$ as was shown for  monolayer graphene by  Crassee \emph{et al.} \cite{kuzmenko2011}
The time reversal symmetry can be broken not only by external magnetic field, but also due to electron-electron interactions. 
Such an example for the latter is one of the possible gapped ground states of bilayer graphene, the so-called quantum anomalous Hall (QAH) state 
(for a general discussion of the possible gapped states in bilayer graphene see Ref.~\onlinecite{PhysRevLett.108.186804}). 
Nandkishore and Levitov has recently proposed that this QAH state   
could be observed by measuring the Kerr rotation~\cite{PhysRevLett.107.097402} in bilayer graphene samples. 
As an extension of Ref.~\onlinecite{PhysRevLett.107.097402}  
the optical Hall and longitudinal conductivities of neutral bilayer graphene  were calculated for four additional 
gapped states by Gorbar \textit{et al.}~\cite{PhysRevB.86.075414}.  The measurement of the Kerr (Faraday) angle has also been  used recently 
to study  other time reversal symmetry breaking systems, such as cuprate superconductors~\cite{Kerr_cond_felter:book,PhysRevB.76.212501,PhysRevLett.100.217004,PhysRevB.80.104508}
and topological insulators~\cite{PhysRevLett.108.087403,PhysRevB.86.235133}.

According to the textbook formula~\cite{Kerr_cond_felter:book,PhysRevB.76.212501,PhysRevLett.100.217004,PhysRevB.80.104508}, 
the Kerr angle $\theta_\mathrm{K}$  for light reflected from a conducting half space 
is proportional to the ac Hall conductivity of the conductor: $\theta_\mathrm{K}  \sim \mathrm{Im} \, \sigma_{xy}(\omega)$. 
However, this formula is no longer valid for atomically thin materials since the thickness of the atomic layer is 
much thinner than the optical wavelength. 
For an atomic layer the relationship between the Hall conductivity and Kerr (Faraday) angle 
$\theta_\mathrm{K}$ ($\theta_\mathrm{F}$) can be derived by solving the Maxwell equations on the two sides 
of the atomic layer and matching solutions at the boundary. Such a 
derivation is presented for bilayer graphene in Ref.~\onlinecite{PhysRevLett.107.097402}, for thin films of 
topological insulators by Tse and MacDonald~\cite{PhysRevLett.105.057401,PhysRevB.82.161104,PhysRevB.84.205327}, 
and for thin films of  topological Weyl semimetals by Kargarian \textit{et al.}~\cite{Sci_Rep_12683:cikk}
Such calculations suggest that Kerr and Faraday angle measurements can also be a  useful tool to characterize heterostructures fabricated recently by stacking atomically thin layers of, e.g., 
graphene, boron-nitride and transition metal dichalcogenides~\cite{korn_2053-1583-2-3-034016,morpurgo2015:cikk,withers2015a,withers2015b}. 
This  calls for a flexible and tractable theoretical framework allowing studies of
 magneto-optical properties of these multilayer systems.

To this end  we develop a simple and versatile method to determine the Kerr and Faraday angles in multilayer
systems. 
In our method the rotation angle $\theta$ and the ellipticity $\eta$ of the polarization for the Kerr and Faraday effect 
are calculated from the total transfer matrix of the multilayer structure.  
The total transfer matrix can always be expressed as a product of many individual transfer matrices 
that can be classified into two different types: i)  transfer matrices corresponding 
to the free propagation in dielectric media, 
and ii) transfer matrices of atomically thin layers with given electric conductivity tensor $\boldsymbol\sigma$. 
As we will show below this kind of classification of the possible transfer matrices 
makes the calculation of polarization dependent reflectivity and transmittivity simple and general. 
Our  approach 
can be easily applied to different 
multilayer structures and for an arbitrary angle of incidence of the electromagnetic radiation. 
Below we also present analytical results for Kerr (Faraday) angle which makes easier the interpretation of experimental results. 
One of the important results  of our work is that  the  Kerr (Faraday) angle can be \emph{enhanced} by 
properly designing the substrate for the atomically thin materials.
To demonstrate how powerful our method is we consider the multilayer setup shown in Fig.~\ref{geo_layers:fig}.   
The atomically thin conductor is a  bilayer graphene flake placed on two layers of dielectric media of refractive indices $n_{1}$ and $n_{2}$. 
Here we only consider the QAH state of bilayer graphene for which the Hall-conductivity $\sigma_{xy}(\omega)$ is finite resulting in Kerr and Faraday rotation. 
Moreover, our method to calculate the Kerr and Faraday rotation can be applied to another exotic state 
called `All' state proposed by Zhang \emph{et al.} 
which breaks the chiral symmetry in bilayer graphene~\cite{PhysRevLett.106.156801}.

We note that a related approach  based on the scattering matrix of the nanostructure has 
been used recently to study the effects of metalic surface states in topological insulator thin films
~\cite{PhysRevLett.105.057401,PhysRevB.82.161104,PhysRevB.84.205327,Slab_Kerr_theory_perpend:cikk}. 
We believe, however, that our transfer matrix method is easier to use in complex nanostructures consisting
of several layers with different optical properties. 
Note that the transfer matrix method has been used for non-interacting graphene layers in 
Ref.~\onlinecite{transfer_Zhan_0953-8984-25-21-215301:cikk} to study the transmission and 
reflection, but  the Kerr (Faraday) effect was not considered there. Thus, our work is
a generalization of  Ref.~\onlinecite{transfer_Zhan_0953-8984-25-21-215301:cikk}.

The paper is organized as follows. 
In Sec. II, we derive the two types of transfer matrices relevant in a multilayer structure described above. 
Moreover, using the total transfer matrix the reflection and transmission amplitudes, 
Kerr and Faraday angles and the ellipticity are given. 
In Sec. III, we present  examples for the application of our transfer matrix method, and analytical 
formulas for the Kerr and Faraday angles for several special cases.  
In order to make our work more readable the main steps of the calculation of the conductivity tensor of 
the gapped bilayer  graphene is presented in Appendix~\ref{calc_opt-cond:app}.     
In Sec. \ref{sec:conclusion} we make our conclusions.


\section{Transfer matrix method for calculating the Kerr and Faraday angles}
\label{transfer_matrix:sec}

In this section  we develop a general and convenient method to calculate the Kerr and Faraday angles 
via the transfer matrix of  layered structures 
{consisting of stacks of dielectric materials and  atomically thin conducting layers, such as graphene.}  
In general, the total transfer matrix of {such} a layered structure is a product of two 
types of transfer matrices. 
The first one corresponds to a free propagation in dielectric media and we shall denote 
it by $\mathbf{M}^{\mathrm{free}}$,  the second one that gives the transfer matrix $\mathbf{M}^{b}$ 
for an atomically thin material 
(e.g., graphene) with electric conductivity $\boldsymbol{\sigma}$.  

\begin{figure}[!ht]
\includegraphics[scale=0.3]{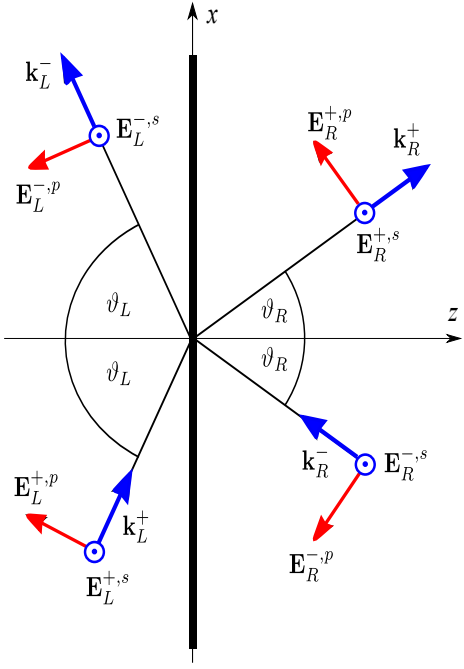}
\caption{\label{transfer:fig}
(Color online) The atomically thin sample is placed on the $x-y$ plane (thick black line). 
The $z$ axis is perpendicular to the interface. 
The figure shows the electric fields of the plane waves at the left (right) side of the interface.
}
\end{figure}

Regarding the geometry, we now consider an atomically thin sample on the $x-y$ plane embedded between 
dielectrics  with refractive indices $n_R$ and $n_L$ at the left and right hand side of the sample, respectively 
as shown in Fig.~\ref{transfer:fig}. 
This figure shows two plane waves with wave vectors $\mathbf{k}_{L}^\pm$ at  the left hand side   
and two plane waves with wave vectors $\mathbf{k}_{R}^\pm$ at the right hand side of the interface.
Here the signs $+/-$ correspond to the direction of the propagation of the waves with respect to the $z$ axis. 
The electric fields of these plane waves at the left and right side of the interface are denoted by subscript $R$ and $L$, respectively. 
The superscripts of these fields are further distinguished by $s/p$ corresponding to the $s/p$ polarized fields, i.e., 
the direction of the field is perpendicular/parallel to the plane of incidence, respectively.  
The transfer matrix $\mathbf{M}^{b}$ connects the electric fields at the left hand side with that of the right hand side of the interface:
\begin{eqnarray}
\label{M4x4_trans:eq}
 \left( \begin{array}{c}
  E^{+, s}_{R} \\[2ex]  E^{+, p}_{R}  \\[2ex] E^{-, s}_{R} \\[2ex] E^{-, p}_{R}  
\end{array}  \right)  
&=& \mathbf{M}^{b}\, \left(
\begin{array}{c}
  E^{+, s}_{L} \\[2ex]  E^{+, p}_{L}  \\[2ex] E^{-, s}_{L} \\[2ex] E^{-, p}_{L}         
\end{array}
\right). 
\end{eqnarray}
In what follows we present our transfer matrix method for the most general case, i.e., for the oblique incidence case. 
From the Maxwell equations one can derive the boundary conditions for the electric and magnetic fields and from that  
the transfer matrix $\mathbf{M}^{b}$ can be extracted. 
Namely, from $\mathrm{rot} \mathbf{E} = - \frac{\partial \mathbf{B}}{\partial t}$, 
$\mathrm{rot} \mathbf{H} = \mathbf{j} + \frac{\partial \mathbf{D}}{\partial t}$ and 
$ \mathbf{j} = \boldsymbol{\sigma} \, \mathbf{E}$  it follows that 
\begin{subequations}%
\label{boundary-cond:eq}%
\begin{align}
\mathbf{\hat{n}}_z \times \left(\mathbf{E}_> - \mathbf{E}_<\right) &= 0 , \\
\mathbf{\hat{n}}_z \times \left(\mathbf{H}_> - \mathbf{H}_<\right) &= \boldsymbol{\sigma} \, \mathbf{E} ,
\end{align}
\end{subequations}%
where $\mathbf{\hat{n}}_z$ is the unit vector along the $z$ axes, $\mathbf{E}_</\mathbf{E}_>$ 
is the electric field at the left/right hand side of the interface. 
The magnetic field of the plane wave in a dielectric is related to the electric field as 
$\mathbf{H} = \sqrt{\frac{\varepsilon_r \varepsilon_0}{\mu_r \mu_0}}\, \frac{\mathbf{k}}{\left|\mathbf{k}\right|} \times \mathbf{E}$. 
For the refractive index $n$ of a dielectric medium we take $n= \sqrt{\varepsilon_r}$ 
since for dielectric the relative permeability constant is $\mu_r \approx 1$.
Now, from Eq.~(\ref{boundary-cond:eq}) we can extract the 4 by 4 transfer matrix $\mathbf{M}^{b}$ 
defined in Eq.~(\ref{M4x4_trans:eq}) and find
\begin{subequations}%
\label{M4x4_2x2:eq}%
\begin{align}
\label{M4x4:eq}
 \mathbf{M}^{b}(n_R,\vartheta_{R}, n_L,\vartheta_{L},\boldsymbol{\sigma}) &= \left(
\begin{array}{cc}
  \mathbf{M}_{11}^{b} & \mathbf{M}_{12}^{b} \\[2ex]
  \mathbf{M}_{21}^{b} & \mathbf{M}_{22}^{b}          
\end{array}
\right),
\end{align}
where
\begin{align}
\mathbf{M}_{11}^{b} &= \frac{1}{2} \left(
\begin{array}{cc}
f_+ -\frac{2\alpha \, \sigma_{yy}}{n_{R}\cos \vartheta_{R}} & 
- \frac{2\alpha \, \sigma_{yx}\, \cos \vartheta_{L}}{n_{R}\cos \vartheta_{R}} \\[2ex]
-\frac{2\alpha \, \sigma_{xy}}{n_{R}} & 
g_+ - \frac{2\alpha \, \sigma_{xx}\, \cos \vartheta_{L}}{n_{R}}
\end{array}
\right), 
\\[2ex]
\mathbf{M}_{12}^{b} &= \frac{1}{2} \left(
\begin{array}{cc}
f_- -\frac{2\alpha \, \sigma_{yy}}{n_{R}\cos \vartheta_{R}} & 
 \frac{2\alpha \, \sigma_{yx}\, \cos \vartheta_{L}}{n_{R}\cos \vartheta_{R}} \\[2ex]
-\frac{2\alpha \, \sigma_{xy}}{n_{R}} & 
g_- + \frac{2\alpha \, \sigma_{xx}\, \cos \vartheta_{L}}{n_{R}}
\end{array}
\right), 
\\[2ex]
\mathbf{M}_{21}^{b} &=  \frac{1}{2} \left(
\begin{array}{cc}
f_- +\frac{2\alpha \, \sigma_{yy}}{n_{R}\cos \vartheta_{R}} & 
 \frac{2\alpha \, \sigma_{yx}\, \cos \vartheta_{L}}{n_{R}\cos \vartheta_{R}} \\[2ex]
-\frac{2\alpha \, \sigma_{xy}}{n_{R}} & 
g_- - \frac{2\alpha \, \sigma_{xx}\, \cos \vartheta_{L}}{n_{R}}
\end{array}
\right), 
\\[2ex]
\mathbf{M}_{22}^{b} &=  \frac{1}{2} \left(
\begin{array}{cc}
f_+ +\frac{2\alpha \, \sigma_{yy}}{n_{R}\cos \vartheta_{R}} & 
- \frac{2\alpha \, \sigma_{yx}\, \cos \vartheta_{L}}{n_{R}\cos \vartheta_{R}} \\[2ex]
-\frac{2\alpha \, \sigma_{xy}}{n_{R}} & 
g_+ + \frac{2\alpha \, \sigma_{xx}\, \cos \vartheta_{L}}{n_{R}}
\end{array}
\right), \\[2ex]
f_\pm &= 1 \pm \frac{n_{L}\cos \vartheta_{L}}{n_{R}\cos \vartheta_{R}} 
\, \, \mathrm{and} \,\, 
g_\pm = \frac{n_{L}}{n_{R}} \pm \frac{\cos \vartheta_{L}}{\cos \vartheta_{R}}, 
\end{align}%
\end{subequations}%
and the angles $\vartheta_{R}$ and $\vartheta_{L}$ 
satisfy the Snell's law: $n_{R} \sin \vartheta_{R} = n_{L} \sin \vartheta_{L}$.
Here the dimensionless conductivity $\boldsymbol{\sigma}$ is in units of $e^2/h$ 
and $\alpha = e^2/(4\pi \varepsilon_0 \hbar c_0) \approx 1/137$ is the fine-structure constant.

One can show that the determinant of the matrix $\mathbf{M}^{b}$ is given by 
\begin{equation}
 \mathrm{det}\, \mathbf{M}^{b} = {\left(\frac{n_L \cos \vartheta_L}{n_R \cos \vartheta_R}\right)}^2.
\end{equation}
Note that it is independent of the conductivity $\boldsymbol{\sigma}$. 

The transfer matrix for free propagation in a dielectric medium is given by
\begin{equation}
\label{Mfree:eq}
  \mathbf{M}^{\mathrm{free}}(d) = \left(
\begin{array}{cccc}
 e^{i k d \cos \vartheta} & 0 & 0 & 0 \\
0 & e^{i k d \cos \vartheta} & 0 & 0 \\
0 & 0 & e^{-i k d \cos \vartheta} & 0 \\
0 & 0 & 0 & e^{-i k d \cos \vartheta} \\
\end{array}
\right). 
\end{equation}
where $k$ is the wave number in the dielectric, $d$ is the thickness of the dielectric medium and $\vartheta$ is the angle 
between the direction of the propagation and the $z$ axes. 
Note that $\mathrm{det}\, \mathbf{M}^{\mathrm{free}} = 1$.

The total transfer matrix is given by the appropriate product of the two building blocks, $\mathbf{M}^{b}$ 
and $\mathbf{M}^{\mathrm{free}}$. 
For example the total transfer matrix for the layered structure shown in Fig.~\ref{geo_layers:fig} reads as
\begin{equation}
\label{Mtotal_geosetup:eq}
\mathbf{M}^{\mathrm{total}} = 
\mathbf{M}^{b}(n_2,n_1,0)\, \mathbf{M}^{\mathrm{free}}(d)\, 
\mathbf{M}^{b}(n_1,n_0,\boldsymbol{\sigma}).
\end{equation}
Here for brevity, we have omitted the dependence of angles $\vartheta_0, \vartheta_1$ and $\vartheta_2$ 
in  two matrices $\mathbf{M}^{b}$. 

The reflection amplitude $\mathbf{r}$ and the transmission amplitude $\mathbf{t}$ 
can be extracted from the total transfer matrix $\mathbf{M}^{\mathrm{total}}$ in the following way.
Consider an incident plane wave which is a superposition of the linear $s$ and $p$ polarized light, 
$\mathbf{E}_i = {(E_i^s,E_i^p)}^T$. 
Now the reflection and transmission amplitudes can be represented by 2 by 2 matrices:
\begin{subequations}
\begin{equation}
\label{r_t:eq}
\mathbf{r} =  \left(
\begin{array}{cc}
 r_{ss} & r_{sp} \\
 r_{ps} & r_{pp}
\end{array}
\right), 
\hspace{5mm} \mathbf{t} =  \left(
\begin{array}{cc}
 t_{ss} & t_{sp} \\
 t_{ps} & t_{pp}
\end{array}
\right),
\end{equation}
and the reflected $\mathbf{r}\mathbf{E}_i$ and the transmitted waves $\mathbf{t}\mathbf{E}_i$ satisfy 
the following equation:  
\begin{eqnarray}
\label{r_t_from_M:eq}
\left(
\begin{array}{c}
 \mathbf{t}\mathbf{E}_i \\
0
\end{array}
\right)
&=& \mathbf{M}^{\mathrm{total}} 
\left(
\begin{array}{c}
  \mathbf{E}_i  \\
  \mathbf{r}\mathbf{E}_i         
\end{array}
\right). 
\end{eqnarray}
\end{subequations}%
Hence, it is easy to obtain
\begin{subequations}
\begin{align}
\label{r_t_M:eq}
\mathbf{r} &=  - {\left(\mathbf{M}_{22} \right)}^{-1} \, \mathbf{M}_{21}, \\
\mathbf{t} &= \mathbf{M}_{11} + \mathbf{M}_{12} \mathbf{r} 
= \mathbf{M}_{11} - \mathbf{M}_{12}\, {\left(\mathbf{M}_{22} \right)}^{-1} \, \mathbf{M}_{21}  \nonumber \\
&= {\left[\left({\mathbf{M}}^{-1}\right)_{11}\right]}^{-1}, 
\end{align}
where the 4 by 4 matrix $\mathbf{M}^{\mathrm{total}}$ is partitioned in the same way as in Eq.~(\ref{M4x4:eq}), i.e., 
\begin{align}
\mathbf{M}^{\mathrm{total}}  &= 
\left(
\begin{array}{cc}
  \mathbf{M}_{11} & \mathbf{M}_{12} \\[2ex]
  \mathbf{M}_{21} & \mathbf{M}_{22}          
\end{array}
\right). 
\end{align}
\end{subequations}
Note that when there is no dissipation, i.e., $\sigma_{xx} = \sigma_{yy}= 0$ and  $\sigma_{xy} = -\sigma_{yx}$  then the unitarity is valid:
\begin{equation}
 \mathbf{r}^+  \mathbf{r} + \frac{n_R \cos \vartheta_R}{n_L \cos \vartheta_L}\, \mathbf{t}^+ \, \mathbf{t} =  \openone, 
\end{equation}
where $ \openone $ is a 2 by 2 unit matrix.
The reflectance $R$ and the transmittance $T$ for  incident light $\mathbf{E}_i$ 
are defined as 
\begin{subequations}
\label{Ref_Trans_intesity:eq}
\begin{align}
 R &= \frac{\mathbf{E}_i^T \, \mathbf{r}^+  \mathbf{r}\, \mathbf{E}_i}{\mathbf{E}_i^T \, \mathbf{E}_i}, \\
 T &= \frac{n_R \cos \vartheta_R}{n_L \cos \vartheta_L}\, 
\frac{\mathbf{E}_i^T \, \mathbf{t}^+  \mathbf{t} \, \mathbf{E}_i}{\mathbf{E}_i^T \, \mathbf{E}_i}.
\end{align}
\end{subequations}
Owing to the dissipation in the atomically thin conductor, some of  the incident light is absorbed, 
and then the absorption $A$ is given by
\begin{equation}
A = 1-R -T. 
\end{equation}

Finally, according to the textbook by Born and Wolf~\cite{Born_Wolf_optics:book} the polarization rotation (Kerr angle) $\theta_\mathrm{K}$ 
and the ellipticity $\eta_\mathrm{K}$ for the reflected wave can be written in the form
\begin{subequations}
\label{theta_eta:eq}
 \begin{align}
\tan (2 \theta_\mathrm{K}) &= \frac{2\, \mathrm{Re}\chi_\mathrm{K}}{1-{\left|\chi_\mathrm{K} \right|}^2}, \label{theta_approx:eq} \\
\sin(2 \eta_\mathrm{K}) &= \frac{2\, \mathrm{Im}\chi_\mathrm{K}}{1+{\left|\chi_\mathrm{K} \right|}^2},
\end{align}
where
\begin{equation}
\label{chi:eq}
 \chi_\mathrm{K}= \begin{cases} 
\frac{r_{ps}}{r_{ss}} & \mbox{for incident linear s-polarized light, } \\[2ex]
-\frac{r_{sp}}{r_{pp}} &  \mbox{for incident linear p-polarized light. } \end{cases}  
\end{equation}
\end{subequations}
For $|\chi_\mathrm{K}| \ll 1$  Eq.~(\ref{theta_approx:eq}) implies that the Kerr angle is given by $\theta_\mathrm{K} \approx \mathrm{Re}\chi_\mathrm{K} $. 
Similar expressions are valid for the polarization rotation $\theta_\mathrm{F}$ (Faraday angle) and the ellipticity $\eta_\mathrm{F}$ 
in the case of transmitted wave, just $\mathbf{r}$ should be replaced by $\mathbf{t}$ in Eq.~(\ref{theta_eta:eq}). 


For dielectrics ($\boldsymbol{\sigma}=0$) our transfer matrix method 
gives the same results as derived, e.g.,  in the classical textbook by Born and Wolf~\cite{Born_Wolf_optics:book}.
If the Hall conductivity $\sigma_{xy}$ is  zero then no polarization rotation emerges, 
i.e., the Kerr and Faraday angles are zeros.
Regarding single and bilayer graphene our method results in the same reflection and transmission amplitudes 
as used by Kuzmenko \textit{et al.}~\cite{PhysRevB.80.165406}


\section{Applications of the transfer matrix method}
\label{Kerr_2layer:sec}

In this section using our general transfer matrix method presented in Sec.~\ref{transfer_matrix:sec} 
we calculate the Kerr rotation angle for the geometrical arrangement shown in Fig.~\ref{geo_layers:fig}. 
To obtain simple analytical results useful for measurements we consider
two special cases here: i) the atomic layer is placed directly on a substrate, i.e., 
the middle dielectric medium with refractive index $n_1$ in Fig.~\ref{geo_layers:fig} is removed. 
ii) the incident light is perpendicular to the plane of the atomic layer. 
Since in our applications the Kerr/Faraday angle is small, i.e., $\theta_\mathrm{K/F}\ll 1 $ 
we use the approximation $\theta_\mathrm{K/F} \approx \mathrm{Re}\chi_\mathrm{K/F} $.

To study numerically the Kerr effect we need to know the frequency dependence of the optical conductivity.
As an example we take the bilayer gapped graphene and calculate its optical conductivity 
using our previously developed method~\cite{Cserti_PhysRevB.82.201405}. 
To make this paper self-contained, in App.~\ref{calc_opt-cond:app} we briefly summarize the main steps to obtain the conductivity. 
We also compare our results with those found in Refs.~\onlinecite{PhysRevLett.107.097402} and~\onlinecite{PhysRevB.86.075414} 
and present some numerical results for bilayer graphene in the QAH state.  


\subsection{Atomic layer on a thick substrate}
\label{thick_substrate:ch}

\begin{figure}
\center
\includegraphics[width=0.45\textwidth]{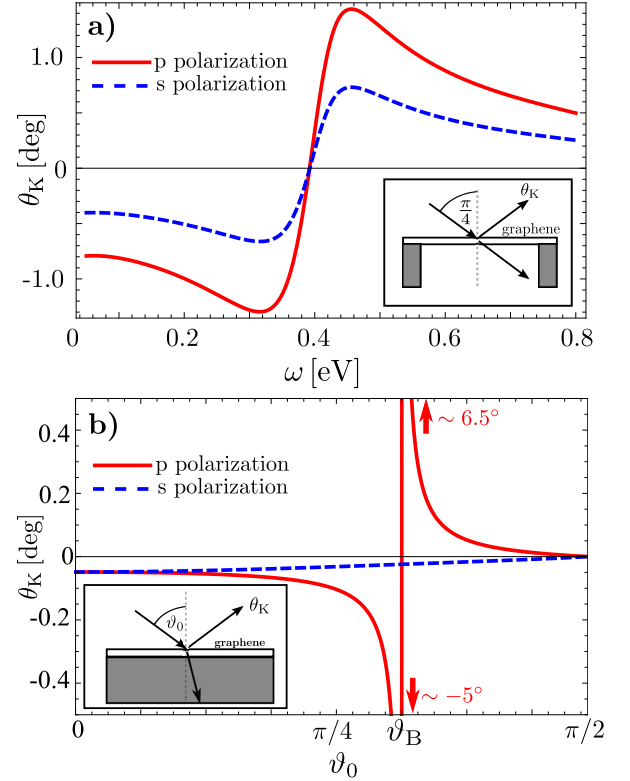}
\caption{\label{Kerr_angle:fig} 
(Color online) a) The Kerr angle for free standing bilayer graphene and incident angle $\vartheta_0 =\pi/4$ as a function of 
the frequency for $s$ and $p$ polarizations. 
b) The Kerr angle in case of bilayer graphene on a thick substrate with refractive index $n_2=1.5$ 
(for geometry see the inset) as a function of the angle of incidence $\vartheta_0$ 
for $p$ (red solid) and $s$ (blue dashed) polarizations at frequency $\omega=0.2$~eV.
The parameters for the calculation of the conductivity: $\gamma_1 = 0.4$~eV, $\eta = 0.05$~eV.
} 
\end{figure}

In this case the total transfer matrix is simply 
$
\mathbf{M}^{\mathrm{total}}=
\mathbf{M}^{b}(n_2,\vartheta_2, n_0,\vartheta_0,\boldsymbol{\sigma})
$
where $\mathbf{M}^{b}(n_2,\vartheta_2, n_0,\vartheta_0,\boldsymbol{\sigma})$ is given by 
Eq.~(\ref{M4x4_2x2:eq}). The Kerr angle is given by 
\begin{subequations}
\label{Kerr_on-substrate:eq}
\begin{equation}
\label{Kerr_on-substrate:eqa}
\theta_\mathrm{K}^{s/p} = -{\rm{Re}}\left[\frac{4\, n_0\, \alpha\, \gamma\, \sigma_{xy}}{a_1^{s/p}
+ 2\, \alpha \, a_2^{s/p}\, \sigma_{xx}+4\, 
\alpha^2\, \gamma\, (\sigma_{xx}^2+\sigma_{xy}^2)}\right],
\end{equation}
where 
\begin{align}
a_1^{s/p} &= (n_2^2-n_0^2)\gamma\pm n_0n_2(1-\gamma^2), \\[2ex]
a_2^{s/p} &= \cos{\vartheta_0} (n_2\mp n_0\gamma)+(n_2\gamma \pm n_0)/\cos{\vartheta_2}, \\[2ex]
\gamma & = \cos{\vartheta_0}/\cos{\vartheta_2}, \hspace{3mm} \mathrm{and} \hspace{3mm} 
n_0\sin{\vartheta_0}=n_2\sin{\vartheta_2}.
\end{align}
\end{subequations}
Here $\vartheta_0$ is the angle of the incident light, and in this subsection the superscript $s$ and the upper sign refer to $s$ polarization, while  
the superscript $p$ and the lower sign refer to $p$ polarization.
At this point it is worth to consider a few special cases of the general formula given by 
Eq.~(\ref{Kerr_on-substrate:eq}).

i) For perpendicular incidence ($\vartheta=0$, $\vartheta_2=0$, $\gamma=1$) the Kerr angle is
given by 
\begin{equation}
\label{Kerr_perp_formula:eq}
 \theta_\mathrm{K} = -{\rm{Re}}\left[\frac{n_0\, \alpha\, \sigma_{xy}}
  {\frac{n_2^2-n_0^2}{4}+\alpha \, n_2 \, \sigma_{xx}+\alpha^2(\sigma_{xx}^2+\sigma_{xy}^2)}\right].
\end{equation}
This result agrees with that derived by Nandkishore and Levitov~\cite{PhysRevLett.107.097402}, 
and Tse and MacDonald~\cite{PhysRevB.84.205327}.
 
ii) For free-standing graphene  ($n_0=n_2=1$, $\vartheta=\vartheta_0=\vartheta_2$, $\gamma=1$) the Kerr angle reads as 
\begin{align}
 \theta_\mathrm{K}^{s/p} &= -{\rm{Re}}\left[\frac{\sigma_{xy}}{ \sigma_{xx}(\cos\vartheta)^{\mp1}+\alpha\, 
 (\sigma_{xx}^2+\sigma_{xy}^2)}\right] \nonumber \\[2ex]
 &\approx - \frac{{\rm{Re}}\left[\sigma_{xy}\right]}{\pi}(\cos\vartheta)^{\pm1}, 
 \label{Kerr_free_stand_formula:eq}
\end{align}
where in the last step we assumed that $\sigma_{xx}$ is approximately equal to $\pi$ in units of $e^2/h$ 
(see, e.g., Ref.~\onlinecite{PhysRevB.77.155409} and our result shown in Fig.~\ref{kerr:fig}a) 
and we neglected the term proportional to $\alpha$ in the denominator. Figure \ref{Kerr_angle:fig}a shows a relatively large Kerr angle plotted as a function of frequency of the incident light for $s$ and $p$ polarization with oblique incidence.

iii) In Eq.~(\ref{Kerr_on-substrate:eqa}) neglecting terms in the denominator that are proportional to $\alpha$ or $\alpha^2$ 
we have 
\begin{equation}
\label{Kerr_neglect_alpha_formula:eq}
\theta_\mathrm{K}^{s/p} \approx \frac{4\, \alpha\,  n_0\, {\rm{Re}}\left[\sigma_{xy}\right]}
{(n_0^2-n_2^2)\left(1\pm\frac{n_0\sin{\vartheta_0}\tan{\vartheta_0}}{\sqrt{n_2^2-n_0^2\sin^2{\vartheta_0}}}\right)}.
\end{equation}

iv) Finally, the Kerr angle for $p$ polarization at the Brewster angle $\vartheta_\mathrm{B}$ 
reads
\begin{align}
\label{Kerr_Brewster_formula:eq}
\theta_\mathrm{K}^{p} &= -{\rm{Re}}\left[\frac{2\, n_0\, \sigma_{xy}}{\sqrt{n_0^2+n_2^2}\, \sigma_{xx}
+2\, \alpha\, (\sigma_{xx}^2+\sigma_{xy}^2)}\right] 
\nonumber \\[2ex]
& \approx \frac{-2\, n_0}{\sqrt{n_0^2+n_2^2}}\, {\rm{Re}}\left[\frac{\sigma_{xy}}{\sigma_{xx}}\right],
\end{align}
where $\vartheta_\mathrm{B} =\arctan{(n_2/n_0)}$ 
(note that now $\gamma=n_0/n_2$). 
In the last step we neglected  the term proportional to $\alpha$. 
For $s$ polarization the Kerr angle is much smaller as can be seen in Fig.~\ref{Kerr_angle:fig}b. 

In what follows we argue that the sensitivity of the detection of the Kerr rotation can be enhanced 
when the incident angle is close to the Brewster angle. 
To see this, we calculated the frequency dependence of the optical conductivity 
for bilayer graphene  assuming that the ground state is the QAH  state. 
(The details of this calculation can be found in Appendix~\ref{calc_opt-cond:app}.)
Using this result we then obtained the Kerr angle as a function of the angle of incidence $\vartheta_0$ 
as shown in Fig.~\ref{Kerr_angle:fig}b. 
As it can be seen  the Kerr angle $\theta_\mathrm{K}$ 
is strongly enhanced for $p$ polarization when  $\vartheta_0=\vartheta_\mathrm{B}$. 
However, using Eq.~(\ref{Ref_Trans_intesity:eq}) 
one can find that at this angle the intensity of the reflected wave significantly drops down.  
Thus, for an optical study of graphene or other atomically thin conducting layers 
the optimal incident angle should be close but not exactly equal to the Brewster angle.


\subsection{Atomic layer on a substrate separated by a dielectric slab: perpendicular incidence}
\label{BLG_Si_SiO2:ch}

Here we will study the multilayer structure shown in Figure \ref{geo_layers:fig}. The total transfer matrix is 
given by Eq.~(\ref{Mtotal_geosetup:eq}) and the Kerr angle reads
\begin{subequations}
 \label{Kerr_Fabry-Perot_1:eq}
\begin{align}
\theta_\mathrm{K} &= {\rm{Re}}\left[\frac{4\, n_0\, a_+^2\, \alpha\, \sigma_{xy}}
{b_+ b_- -4\, a_+^2\, \alpha^2\, \sigma_{xy}^2}\right],  \,\,\, \mathrm{where} \\ 
a_{\pm} &=(n_1-n_2)e^{2i k d}\pm(n_1+n_2), \\
b_\pm &= a_+ (n_0 \pm 2\, \alpha\, \sigma_{xx}) \mp a_-n_1,
\end{align}
\end{subequations}
and $k=\omega \, n_1/c$ is the wave number in the dielectric with refractive index $n_1$ and $\omega$ is the 
frequency of the incident light. Here (in contrast to Sec. \ref{thick_substrate:ch}) the upper/lower signs are only introduced to make the expressions more compact.

\begin{figure} 
\includegraphics[scale=0.65]{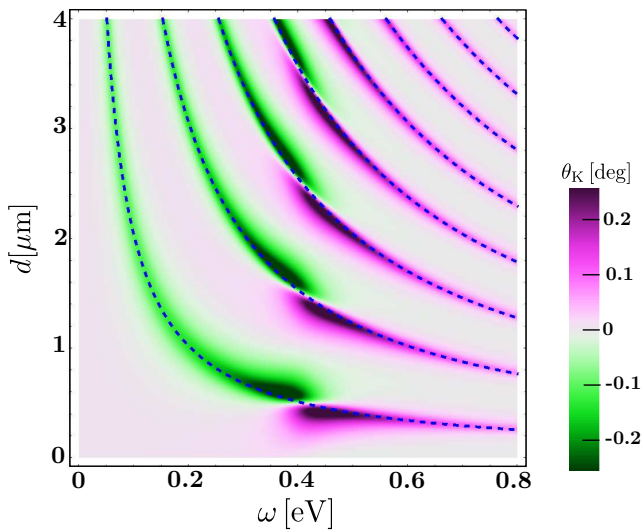}
\caption{\label{Kerr_BLG_SiO_2_Si_omega:fig} 
(Color online) The Kerr angle for a bilayer graphene sheet with a multilayer structure shown in  Fig.~\ref{geo_layers:fig} 
as a function of the frequency and the thickness $d$ of SiO$_2$ layer at perpendicular incidence. 
The dashed lines show the resonance conditions derived analytically in Eq.~(\ref{eq:d-omega-2}). 
The parameters: $\gamma_1 = 0.4$~eV, $\eta = 0.05$~eV, $n_1=1.5$, $n_2=3.5$.
} 
\end{figure}

We now argue that  in this setup  an appropriate choice of substrate thickness $d$ makes the detection of $\theta_\mathrm{K}$ easier in a somewhat similar way as in monolayer graphene flakes where the visibility is enhanced ~\cite{korn_2053-1583-2-3-034016,morpurgo2015:cikk}.  
We again consider only the QAH state of bilayer graphene and calculate 
the dependence of the Kerr angle on the frequency $\omega$ and the thickness $d$ of the SiO$_2$ dielectric. 
The substrate is made of  Si and the electromagnetic wave incident perpendicular to the interface comes from vacuum 
($n_0=1$).  The optical Hall conductivity of the bilayer graphene is calculated at  zero chemical potential 
and temperature (see  Appendix~\ref{calc_opt-cond:app} for details).
The results for Kerr angle $\theta_\mathrm{K}$ are shown in Fig.~\ref{Kerr_BLG_SiO_2_Si_omega:fig}. 
One can see from Fig.~\ref{Kerr_BLG_SiO_2_Si_omega:fig} that the Kerr angle is enhanced 
along certain lines on the $d-\omega$ plane. 
This is a consequence of the Fabry-Perot type resonance. 
Indeed, from Eq.~(\ref{Kerr_Fabry-Perot_1:eq}) we can derive an approximate analytical expression 
for the resonance condition, which is given by $b_+ b_- = 0$, i.e.,  the first term in the 
denominator vanishes. Note, that the second term in the denominator is proportional to the 
square of the fine-structure constant and therefore it is generally a small term.
Using the definitions of $b_\pm$ given by Eq.~(\ref{Kerr_Fabry-Perot_1:eq}) the condition $b_+ b_-=0$  
leads to $n_0^2a_+^2-n_1^2a_-^2=0$. 
This equation can be satisfied in two cases: 
\begin{subequations}
\label{d-omega:eq}
 \begin{align}
  d &= \frac{c\, \pi}{\omega n_1}\, N , \,\,\, \mathrm{if} \,\,\,  n_2=n_0, \\
  d &= \frac{c\, \pi}{\omega n_1}\left(N +\frac{1}{2}\right), \,\,\, \mathrm{if} \,\,\,  n_1=\sqrt{n_0n_2}, \label{eq:d-omega-2}
 \end{align}
\end{subequations}
where $N$ is an integer. For SiO$_2$ layer ($n_1=1.5$, see Ref.~\onlinecite{refractive_index_SiO2_Malitson:cikk}) and Si substrate 
($n_2=3.5$, see Ref.~\onlinecite{PhysRevB.27.985}) the above condition $n_1=\sqrt{n_0 \, n_2}$ cannot be satisfied perfectly.
Nevertheless, it is clearly seen in Fig.~\ref{Kerr_BLG_SiO_2_Si_omega:fig} that $\theta_\mathrm{K}$  is strongly enhanced 
along lines where   Eq.~(\ref{eq:d-omega-2}) is  approximately satisfied.

As a brief summary of our findings in Secs.~\ref{thick_substrate:ch} and \ref{BLG_Si_SiO2:ch} regarding the Kerr angle, the following conclusions can be drawn:

i) According to Eq.~(\ref{Kerr_free_stand_formula:eq}) the real part of the Hall conductivity $\sigma_{xy}$ 
for free-standing graphene can directly be determined by measuring the relatively large Kerr angle. 

ii) From Eq.~(\ref{Kerr_Brewster_formula:eq}) it follows that the Kerr angle can be enhanced when the atomically 
thin material is placed on a bare substrate 
and the incident angle $\vartheta_0$ of the light is close to the Brewster angle $\vartheta_\mathrm{B}$. 

iii) If the atomically thin material and the substrate are separated by a dielectric slab 
then owing to a Fabry-Perot type resonance the Kerr angle can  be enhanced 
if the frequency of the incident light is tuned according to Eq.~(\ref{d-omega:eq}).


\subsection{Faraday effect for atomic layer on a thick substrate}
\label{Faraday_thick_substrate:ch}

In this section we consider the same multilayer structure as in shown Fig.~\ref{geo_layers:fig}. 
 except that the dielectric medium with refractive index $n_2$ is replaced by vacuum. 
We consider that  the incident light coming from the vacuum is perpendicular to the 
conducting sheet.  
Using the theory outlined in Sec.~\ref{transfer_matrix:sec} one can derive the following simple analytical expression 
for the Faraday angle $\theta_\mathrm{F}$. 
\begin{subequations}
 \label{Faraday:eq}
\begin{align}
\theta_\mathrm{F} &= -{\rm{Re}}\left[\frac{\sigma_{xy}\left(a_+ +a_- \right)}{b_+ - b_-} \right], \,\,\,  \mathrm{where} \\
a_\pm &=  e^{\pm i k d} \left(n_1\mp n_0\right), \\ 
b_{\pm} &= e^{\pm i k d} \left(n_0\mp n_1\right)\left(n_0 \mp n_1 + \sigma_{xx}\right),
\end{align}
\end{subequations}
and $k=\omega \, n_1/c$ is the wave number in the dielectric with refractive index $n_1$ 
and $\omega$ is the frequency of the incident light. Here (in contrast to Sec. \ref{thick_substrate:ch}) the upper/lower signs are only introduced to make the expressions more compact.

\begin{figure} 
\includegraphics[scale=0.65]{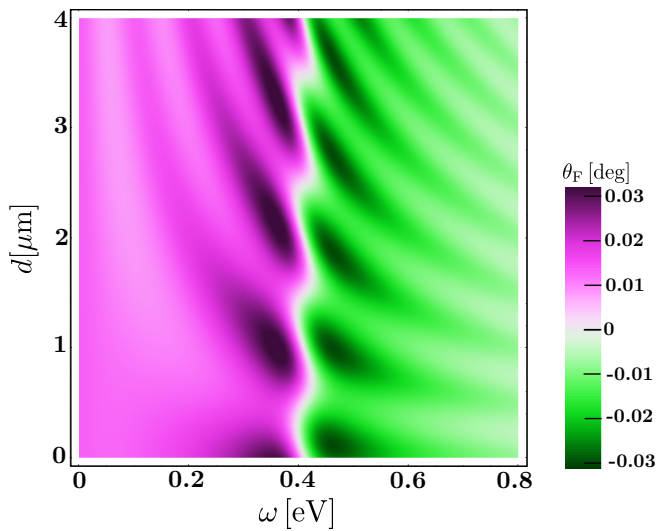}
\caption{\label{Faraday_BLG_SiO_2_Si_omega:fig} 
(Color online) The Faraday angle as a function of $\omega$ and the thickness $d$ 
for a bilayer graphene sheet placed on a substrate of refractive index $n_1 = 1.5$ and at perpendicular incidence. 
The parameters: $\gamma_1 = 0.4$~eV, $\eta = 0.05$~eV, $n_0=1$.
} 
\end{figure}

As in Sections \ref{thick_substrate:ch} and \ref{BLG_Si_SiO2:ch} for our numerical calculations 
we take bilayer graphene in QAH state.
Figure~\ref{Faraday_BLG_SiO_2_Si_omega:fig} shows the Faraday angle $\theta_\mathrm{F}$ as a function of the frequency $\omega$ and the thickness $d$ of the substrate for perpendicular incidence.
The enhancement of the Faraday angle that can be seen in Fig.~\ref{Faraday_BLG_SiO_2_Si_omega:fig} is
consequence of the  local extrema of  $\sigma_{xy}$ as a function of $\omega$ 
(see Fig.~\ref{kerr:fig}b in Appendix \ref{calc_opt-cond:app}).
However, this angle is still smaller by one order of magnitude than the maximum values of the Kerr angle shown 
in Fig.~\ref{Kerr_BLG_SiO_2_Si_omega:fig}. 
Thus, measuring the Kerr angle is more suitable than the Faraday angle to explore whether the time reversal symmetry 
is broken or not in bilayer graphene. 


\section{Conclusions}
\label{sec:conclusion}

In this work we developed a general and versatile approach to calculate the rotation of the 
polarization of reflected and transmitted light (Kerr and Faraday effects) that is  
incident on multilayer systems consisting of atomically thin conducting layers and dielectrics. 
Introducing two kinds of transfer matrices as building blocks provides a powerful method to determine the transfer matrix of such multilayers in a simple and systematic way.
From the transfer matrix we presented expressions for the intensity of the reflected and transmitted light, 
and the rotation angle and ellipticity of the light polarization. 
The expressions of these quantities are also applicable for oblique incidence of light.
As an example we considered a geometrical arrangement of the multilayers as shown in Fig.~\ref{geo_layers:fig} 
and for several special cases we derived analytical results for the Kerr angle. 
In particular, we found that  if the angle of incidence is close to the Brewster angle
the Kerr angle is enhanced allowing  easier detection. 
We would like to emphasize that these analytic results can be applied to any 2D conducting materials layered with dielectrics. 

In our numerical calculations the atomically thin conducting layer is taken to be a bilayer graphene using a four-band model.
 The measurement of the Kerr and/or Faraday rotation provides a simple optical 
method to determine whether the ground state is the quantum anomalous Hall state characterized by spontaneously 
broken time-reversal symmetry or not~\cite{PhysRevLett.107.097402,PhysRevB.86.075414}. 
Our newly developed transfer matrix method is an efficient procedure to design such multilayer structures in which the Kerr angle can be enhanced.  
As an example we showed that the Kerr angle can be maximized by tuning the thickness of the SiO$_2$ layer.

We believe that our work for calculating the Kerr and Faraday rotations can be applied 
to interpret and design experiments on complex multilayers consisting of atomically 
thin conducting materials and dielectrics.

\acknowledgements 

We would like to thank  L. Oroszl\'any, A. P\'alyi and L. Tapaszt\'o for helpful discussions.
This work is supported by the National Research, Development and Innovation Office under the contracts No.~K108676.

\appendix

\section{Calculation of the optical conductivity for gapped bilayer graphene}
\label{calc_opt-cond:app}

To calculate the optical conductivity of any 2D material we applied our general method developed earlier in Ref.~\onlinecite{Cserti_PhysRevB.82.201405}.
In this approach we start with an arbitrary multi band system
described by a matrix Hamiltonian in a Bloch wavefunction basis:
$H_{ab}(\textbf{k})$, where $a,b= 1,2, \cdots N$ are the band indices (here $N$ is the number of bands of the system).
Here each matrix element $H_{ab}(\textbf{k})$ is a differentiable function of the wave number $\textbf{k}$ corresponding 
to the Bloch states.

As an example we take the same four-band model of gapped bilayer graphene that is used by Gorbar \textit{et al.} in Ref.~\onlinecite{PhysRevB.86.075414}. 
This was an extension of the two-band model used by Nandkishore \textit{et al.} in Ref.~\onlinecite{PhysRevLett.107.097402} 
to describe the broken symmetry in bilayer graphene at low energy. 
The 4 by 4 Hamiltonian is given by
\begin{equation}
\label{bilayerH2c:eq}
H = \xi \left( \begin{array}{cccc}
\Delta_{\xi s} & 0 & 0&  \hbar v_F k_{-}\\
0 & -\Delta_{\xi s} &\hbar v_F k_{+} & 0\\
0 &\hbar v_F k_{-} & 0 & \xi \gamma_1\\
\hbar v_F k_{+}& 0 & \xi \gamma_1 &0
\end{array} \right),
\end{equation}
where $k_{\pm}=k_x\pm i k_y$, and $\xi = \pm 1$ and $s=\pm 1$ are valley and spin quantum numbers, respectively,
while $v_F \approx 10^6$~m/s is the Fermi velocity and 
$\gamma_1 = 0.38$ eV is the strongest interlayer hopping.
Here the most general gap reads as 
\begin{equation}
 \Delta_{\xi s} = U + s U_T + \xi \Delta_T + \xi s \Delta,
\end{equation}
where  $U$, $U_T$, $\Delta$ and $\Delta_T$  are constants related to different gapped ground states.

The four eigenvalues of the Hamiltonian (\ref{bilayerH2c:eq}) are 
$ E_{1,2} (k)=E_{\pm}(k)$ and $ E_{3,4} (k)=-E_{2,1}(k)$, where 
\begin{equation}
\label{eigenvalues:eq}
E_{\pm}^2 = x+\frac{\Delta_{\xi s}^2+\gamma_1^2}{2}  
\pm\sqrt{\frac{(\gamma_1^2-\Delta_{\xi s}^2)^2}{4}+{(\gamma_1^2+\Delta_{\xi s}^2) x}},
\end{equation}
while $x=(\hbar v_F k)^2$ and $k$ is the magnitude of the wave vector $\mathbf{k} = (k_x,k_y)$.

In general the complex optical conductivity $\sigma_{ij}(\omega)$ can be calculated from the 
current-current correlation function $\Pi_{i j}(i\nu_m) $ 
using the usual analytic continuation~\cite{Mahan_book}
$i \nu_m \to \hbar\omega + i \eta$, and it is given by 
\begin{equation}
 \label{cond_def:eq}
\sigma_{ij}(\omega)=\frac{i e^2}{\hbar^2\omega}\, \Pi_{ij}(i\nu_m \to \hbar\omega +i \eta), 
\end{equation}
where $i,j = x,y$ and $\eta$ is the inverse lifetime of the particle. 
To calculate the current-current correlation function we applied our general method developed earlier 
in Ref.~\onlinecite{Cserti_PhysRevB.82.201405} in the usual bubble approximation. 
To this end it is useful to write the Hamiltonian as $H  = \sum_a E_a Q_a$, where $Q_a = | a \rangle \langle a |$  
are the projector operators, 
and  $E_a$ and  $ |a \rangle $ are the eigenenergies and the corresponding eigenvectors of the Hamiltonian $H$, 
and in our case $a=1,2,3,4$. 
The projectors $Q_a$ satisfy the usual relation $Q_a Q_b = \delta_{a b} \, Q_a $. 
Then the current-current correlation function $\Pi_{i j}(i\nu_m) $ with current operator
$\textbf{J} = \frac{\partial H}{\partial \textbf{k}}$
(in units of $e/\hbar$ which is taken into account in the expression of the conductivity) reads
\begin{subequations}
\begin{align} 
\label{PIJ-Ham-def:eq}
  \hspace{-5mm}  \Pi_{i j}(i\nu_m)  &= \frac{1}{V}\, \sum_{\textbf{k}} \sum_{a,b} K_{ba}(i\nu_m) \,
   \textrm{Tr} \Bigl(\, \frac{\partial H}{\partial k_i}\, Q_a\, \frac{\partial H}{\partial k_j}\, Q_b\,  \Bigr),  \\
    K_{ab}(i\nu_m)  &=  \frac{n_F(E_a -\mu) - n_F(E_b -\mu) }{i\nu_m + E_a -E_b},
\end{align}
\end{subequations}
where $n_F(E) = 1/(e^{\beta E}+1)$ is the usual Fermi distribution and the trace is taken over the band indices. 
Note that to calculate the function $K_{ab}(i\nu_m) $ we have used the usual summation techniques over the 
Matsubara's frequencies~\cite{Mahan_book}.
Here we would like to emphasize that the projector operators
$Q_a$ can be calculated without knowing the eigenvectors $ | a \rangle $ of the Hamiltonian $H$.
Indeed, let $H$ be an $N \times N$ hermitian matrix with $s\leq N$ distinct eigenvalues, $E_a, \dots, E_s$, and
then the matrix $H$ can be decomposed in terms of projector matrices as $H=\sum_a E_a Q_a$, where
the projector matrix $Q_a$ for $a = 1,\dots, s$ (in the mathematical literature called 
Frobenius covariant~\cite{Topics_in_Matrix:book}) is given by
\begin{eqnarray}
\label{Rozsa_tetel:eq}
Q_a &=& \prod_{\substack{b=1 \\ b\ne a}}^s \frac{H- E_b\, I_N }{E_a-E_b},
\end{eqnarray}%
where $I_N$ is the $N \times N$ unit matrix. 
The proof of (\ref{Rozsa_tetel:eq}) is based on the Cayley–-Hamilton theorem~\cite{Topics_in_Matrix:book,Lax_Peter:book}.
This theorem greatly simplifies the calculation of the current-current correlation function both analytically and numerically. 
Moreover, one can avoid to evaluate the spectral function of the Green's function 
used for example by Nicol and Carbotte in Ref.~\onlinecite{PhysRevB.77.155409}. 
 
In particular, for Hamiltonian (\ref{bilayerH2c:eq}) we find the correlation function for chemical 
potential $\mu =0$ and at zero temperature
\begin{subequations}
\label{Pxx_Pxy:eq}
\begin{align}
\Pi_{ij}(i\nu_m) &= \sum_{\xi=\pm 1,s=\pm 1}\,\frac{1}{8\pi}\int_0^\infty \, 
dx\left(\frac{2Z_{ij}^{13}}{i\nu_m+E_++E_-} \right.\nonumber \\
&   -\frac{2Z_{ij}^{31}}{i\nu_m-E_+-E_-} \nonumber +\frac{Z_{ij}^{14}}{i\nu_m+2E_+}-\frac{Z_{ij}^{41}}{i\nu_m-2E_+} \\
& \left.+\frac{Z_{ij}^{23}}{i\nu_m+2E_-}-\frac{Z_{ij}^{32}}{i\nu_m-2E_-}\right),
\end{align}
where we introduced a notation for $Z_{ij}^{ab}$:
\begin{equation}
Z_{ij}^{ab}=\frac{1}{\pi\hbar^2v_F^2}\int_0^{2\pi} d\varphi \,\textrm{Tr}\left[
\frac{\partial H}{\partial k_i}Q_a\frac{\partial H}{\partial k_j}Q_b\right],
\end{equation}
\end{subequations}
and the integration is with respect to the polar angle $\varphi$ of the wave vector 
$\mathbf{k} = k (\cos \varphi, \sin \varphi)$. 
Since the expressions for the projectors $Q_a$ are very lengthy we do not present them here. 
However, after taking the trace and performing the integration the expressions for $Z_{ij}^{ab}$
are greatly simplified and here we list only the relevant $Z_{ij}^{ab}$ 
\begin{subequations}
\label{Trab:eq}
\begin{align}
Z_{xy}^{13} &=Z_{xy}^{24}=-Z_{xy}^{31} = -Z_{xy}^{42} \nonumber \\
& =\frac{i\xi\Delta_{\xi s}(\Delta_{\xi s}^2-\gamma_1^2)(\gamma_1^2-x-E_+E_-)}{E_+ E_-(E_+-E_-)^2(E_++E_-)},\\
Z_{xy}^{23} &= -Z_{xy}^{32} = -\frac{8i\gamma_1^2\xi\Delta_{\xi s} x}{E_-(E_+^2-E_-^2)^2},\\
Z_{xy}^{14} &= -Z_{xy}^{41} = -\frac{8i\gamma_1^2\xi\Delta_{\xi s}x}{E_+(E_+^2-E_-^2)^2}, \\
Z_{xx}^{13} &=Z_{xy}^{24}=Z_{xx}^{31} = Z_{xx}^{42}\nonumber \\
& =\frac{\left(\gamma_1^2- \Delta_{\xi s}^2\right)^2E_+E_-+x\Delta_{\xi s}^2\left[\left(E_++E_-\right)^2-4\gamma_1^2\right]}{E_+ E_- {\left( E_+^2-E_-^2\right)}^2},\\
Z_{xx}^{23} &= Z_{xx}^{32}  = \frac{4\gamma_1^2(E_-^2+\Delta_{\xi s}^2) x}{E_-^2(E_+^2-E_-^2)^2},\\
Z_{xx}^{14} &= Z_{xx}^{41} = \frac{4\gamma_1^2(E_+^2+\Delta_{\xi s}^2) x}{E_+^2(E_+^2-E_-^2)^2}.
\end{align}
\end{subequations}
Now inserting Eqs.~(\ref{Trab:eq}) into (\ref{Pxx_Pxy:eq})  
we find an analytical form for the current-current correlation function $\Pi_{ij}$ at zero temperature. 
Then using Eqs.~(\ref{cond_def:eq}) we obtain the complex conductivity:
\begin{widetext}
\begin{subequations}
\label{full_cond_bilayer:eq}
\begin{align}
\label{sigma_xx:eq}
\sigma_{xx}(\omega) &=\frac{e^2}{h} \sum_{\xi=\pm 1,s=\pm 1}\, \frac{1}{i\hbar\omega} \int_0^\infty dx
\left\{
\frac{\left(\gamma_1^2- \Delta_{\xi s}^2\right)^2E_+E_-+x\Delta_{\xi s}^2\left[\left(E_++E_-\right)^2-4\gamma_1^2\right]}{E_+ E_- {\left( E_+-E_-\right)}^2\left(E_++E_-\right)}\, 
\left(\frac{1}{\left(E_++E_-\right)^2-\left(\hbar\omega+i\eta\right)^2}\right) \right.   \nonumber \\
&+\left. 
\frac{4x\gamma_1^2}{{\left( E_+^2-E_-^2\right)}^2}\, \left[ 
\frac{E_+^2+\Delta_{\xi s}^2}{E_+}\, \left(\frac{1}{4E_+^2-\left(\hbar\omega+i\eta\right)^2}\right)
+\frac{E_-^2+\Delta_{\xi s}^2}{E_-}\, \left(\frac{1}{4E_-^2-\left(\hbar\omega+i\eta\right)^2}\right) \right] \right\},  \\[2ex]
\label{sigma_xy:eq}
\sigma_{xy}(\omega) &= -\frac{e^2}{h} \, \sum_{\xi=\pm 1,s=\pm 1}\, \frac{(\hbar\omega+i\eta) \xi \Delta_{\xi s} }{\hbar\omega} \int_0^\infty dx
\left\{
\frac{\left(\gamma_1^2- \Delta_{\xi s}^2\right)\left(\gamma_1^2- x - E_+ E_-\right)}{E_+ E_- {\left( E_+-E_-\right)}^2\left(E_++E_-\right)}\, 
\left(\frac{1}{\left(E_++E_-\right)^2-\left(\hbar\omega+i\eta\right)^2}\right) \right.  \nonumber \\
&+\left. 
\frac{4x\gamma_1^2}{{\left( E_+^2-E_-^2\right)}^2}\, \left[ 
\frac{1}{E_+}\, \left(\frac{1}{4E_+^2-\left(\hbar\omega+i\eta\right)^2}\right)
+\frac{1}{E_-}\, \left(\frac{1}{4E_-^2-\left(\hbar\omega+i\eta\right)^2}\right) \right] \right\}, 
\end{align}
\end{subequations}
\end{widetext}
while $\sigma_{yy}(\omega)=\sigma_{xx}(\omega)$ and $\sigma_{yx}(\omega)=-\sigma_{xy}(\omega)$. 

\begin{figure} [h!]
\includegraphics[scale=0.65]{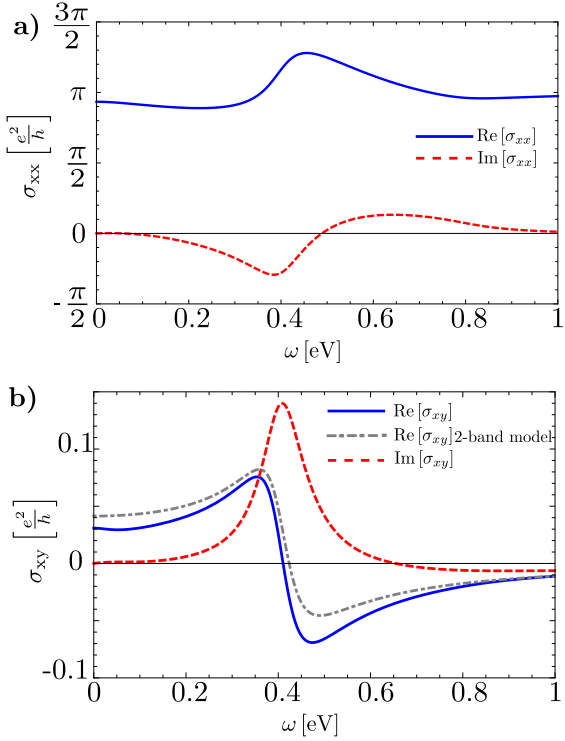}
\caption{\label{kerr:fig} 
(Color online) a) The real (blue solid line) and imaginary part (red dashed line) of 
the longitudinal optical conductivity $\sigma_{xx}$ (in units of $e^2/h$) 
given by Eq.~(\ref{sigma_xx:eq}). 
b) The real (blue solid line) and imaginary part (red dashed line) of 
the optical Hall conductivity $\sigma_{xy}$ (in units of $e^2/h$) calculated from Eq.~(\ref{sigma_xy:eq}), 
and the real part of $\sigma_{xy}$ (gray dash-dot line) calculated from the two-band model 
according to Ref.~\onlinecite{PhysRevLett.107.097402}. 
In both panels the chemical potential and temperature are zero, 
and the parameters are $\Delta_T = 1$~meV, $\gamma_1 = 0.4$~eV, $\eta = 0.05$~eV.
(Note that our parameter $\eta = 2\Gamma$ used in Ref.~\onlinecite{PhysRevB.86.075414}).
} 
\end{figure}

At this point the above form of the conductivity tensor is valid for arbitrary gap parameters $U$, $U_T$,  $\Delta$ and $\Delta_T$.
From now on we take $U=U_T=\Delta=0$ and  for $\Delta_T$ we use the same value as in Ref.~\onlinecite{PhysRevB.86.075414}.
We plotted
the real and imaginary part of the complex longitudinal optical conductivity $\sigma_{xx}$ 
given by Eq.~(\ref{sigma_xx:eq}) (see Fig.~\ref{kerr:fig}a), 
and the real and imaginary part of the complex optical Hall-conductivity $\sigma_{xy}$ 
calculated from Eq.~(\ref{sigma_xy:eq}) (see Fig.~\ref{kerr:fig}b).
Furthermore, we also compare our result with that obtained 
by Nandkishore and Levitov using the simplified two-band model for bilayer graphene~\cite{PhysRevLett.107.097402} 
(see the gray dash-dot line in Fig.~\ref{kerr:fig}b). 
As can be seen from Fig.~\ref{kerr:fig}b the result from the two-band model agrees well with our four-band calculations.

Note that the current-current correlation function obtained from Eq.~(\ref{Pxx_Pxy:eq}) 
agrees exactly with that obtained by Gorbar \textit{et al.} using a different method~\cite{PhysRevB.86.075414}. 
However, the conductivity in Eq.~(\ref{full_cond_bilayer:eq}) differs from that given in Ref.~\onlinecite{PhysRevB.86.075414} by a factor 
$(\omega+i\eta)/\omega $. 
As can be shown numerically this analytic difference is relevant only at low frequencies, 
namely for $\omega \lessapprox \eta$. 

Note that as a check of our calculation of the optical conductivity it can be shown that
\begin{equation}
 \lim_{\Delta_T\rightarrow0}\lim_{\omega\rightarrow0}\lim_{\eta\rightarrow0}\textrm{Re}[\sigma_{xy}(\omega)]=\frac{4e^2}{h}
\end{equation}
when the spin and valley degeneracy are taken into account.


%

\end{document}